\documentclass[aps,prb,%
twocolumn,%
groupedaddress,amsmath,amssymb]{revtex4}

\usepackage{amssymb}
\usepackage{amsmath}
\usepackage{graphicx}
\usepackage{subfigure}
\usepackage{textcomp}
\usepackage{color}
\usepackage{amsfonts}
\usepackage{bbold}
\usepackage{dsfont}
\usepackage{epsfig}
\usepackage{hyperref}

\newcommand{\arctanh}[1]{\operatorname{arctan}}

\begin{document}

\title{Silicon-based molecular switch junctions} 
\date{\today} 
\author{Daijiro Nozaki
and Gianaurelio Cuniberti}
\email{g.cuniberti@tu-dresden.de}
\affiliation{Institute for Materials Science and Max Bergmann Center of Biomaterials, 
Dresden University of Technology, D-01069 Dresden, Germany
}

\begin{abstract}
In contrast to the static operations of conventional semiconductor devices, the dynamic conformational freedom in molecular devices opens up the possibility of using molecules as new types of devices such as a molecular conformational switch or for molecular data storage. Bistable molecules, with \textit{e.g.}~two stable \textit{cis} and \textit{trans} isomeric configurations, could provide, once clamped between two electrodes, a switching phenomenon in the nonequilibrium current response. Here, we model molecular switch junctions formed at silicon contacts and demonstrate the potential of  tunable molecular switches in electrode/molecule/electrode configurations. Using the non equilibrium Green function approach implemented with the density-functional-based tight-binding theory, a series of properties such as electron transmissions,  \textit{I-V} characteristics  in the different isomer-conformations, and potential energy surfaces as a function of the reaction coordinates along the \textit{trans} to \textit{cis} transition were calculated. Furthermore, in order to investigate stability of molecular switches in ambient conditions, molecular dynamics (MD) simulations at room temperature were performed and time-dependent fluctuations of the conductance along the MD pathways were calculated. Our numerical results show that the transmission spectra of the \textit{cis} isomers are more conductive than \textit{trans} counterparts inside the bias window for all two model molecules. The  current-voltage characteristics consequently show the same trends. Additionally, the calculations of time-dependent transmission fluctuations along the MD pathways have shown that the transmission in \textit{cis} isomers is always significantly larger than that of \textit{trans} counterparts showing that molecular switches can be expected to work as robust molecular switching components.  
\end{abstract}


\maketitle

\section{Introduction}
Since the successful measurement of current flow through individual molecules in the past few decades, various kinds of molecular devices have been demonstrated by many groups, showing interesting properties such as negative differential resistance\cite{Reed} and memory effects.\cite{1Gio-Book05} The realization of electronic function at the molecular scale, visionally proposed by Aviram and Ratner already in 1974,\cite{Aviram74} is gradually finding experimental support due to the development of scanning probe technologies.\cite{Frei-Berlin} Nevertheless, many challenges are still ahead. The conductance of a molecule connected between contacts changes dramatically depending on many parameters such as the coupling strength between the molecule and of the contacts, the electronic structure of the molecule and contacts, and molecular conformational changes. Therefore, it is highly desirable to understand the relationship between those influential parameters and transport properties for the design of controllable molecular devices. 

In contrast to the static operations of conventional semiconductor devices, the dynamic conformational freedom in molecular devices opens the possibility of using individual molecules as a new type of device such as a conformational molecular switch or for molecular data storage. The ability to control the conductance at the single molecule scale will have a great impact in nanoscale electronics as well as other fields such as chemical sensors\cite{Cui01} or medical diagnostics.\cite{Patolsky04} Photochromic organic molecules having bistable characteristics are attractive materials for the design of the molecular switches since a great variety of properties such as their molecular conformations and their electronic structures can be reversibly altered from one state to another by external stimulations such as light irradiation and current-pulses. Azobenzene, diarylethene and their derivatives are representative photochromic molecules.
Recently there have been many theoretical and experimental studies on molecular switches.\cite{Feringa01} Theoretical studies have explored the use of azobenzene as a molecular switch when connected by Au electrodes\cite{Zhang04} or carbon nanotubes (CNTs).\cite{Valle07}  Promising results were obtained, as a significant change in conductance was seen between the two conformations. Additionally, the study of diarylethene derivatives connected to Au contacts also have presented large change in conductance between open or closed forms.\cite{Staykov07}

Several experimental groups have addressed the fabrication of molecular switches consisting of molecules with bistable characteristics sandwiched between contacts. Dulic \textit{et al.}~have created diarylethene derivative-based molecular switch junctions connected between gold contacts by means of the mechanically controlled break junction (MCBJ) technique and measured the current flow through the junction before and after UV irradiation.\cite{Dulic03} The current flow through the molecule was suddenly increased by the UV irradiation. Additionally, the current flow did not decay after stopping the irradiation. Although the reversible switching has not been observed, conformational switching and the control of current flow through the single molecular device have been realized.

L{\"o}rtscher~\textit{et al.}~have fabricated molecular junctions where single bipyridyl-dinitro oligophenylene dithiole molecules have been contacted between gold contacts with the MCBJ technique.\cite{IBM06} They have demonstrated reversible molecular switching by applying current pulses across the gold contacts. 
Additionally, Henzl \textit{et al.}\cite{Henzl06}~and Choi \textit{et al.}\cite{Choi}~have reported reversible \textit{cis-trans} isomerization of a single azobenzene derivative at the Au(111) surfaces induced by the inelastic excitation of vibrational states by an STM tip. The experiment by Alemani \textit{et al.} have reported mechanical reversible isomerization of a single azobenzene derivative on the Au(111) induced by an electric field applied from an STM tip not in contact with the molecule.\cite{Alemani06}
 

In this article, we explore the potential of azobenzene-based molecules  as reversibly tunable molecular switches in electrode/molecule/electrode configurations. We have modeled two kinds of molecular switch junctions formed at silicon contacts. In order to improve the molecule-contact coupling the molecules are covalently attached to Si contacts.  We present a detailed analysis of the energetics, potential energy surface (PES) as a function of the reaction coordinate along the \textit{trans} to \textit{cis} transition, the stability along MD pathways, and transport properties in the two conformations using NEGF approach based on DFTB method.

 This paper is organized as follows. In Sec.~II, we present computational details which are used in our study. In Sec.~III, we show the molecular structures, electronic structures, and PESs of molecular switches in gas phase and connected between Si contacts. In Sec.~IV, we show the transport properties and the stability of the molecules between Si contacts. We conclude in Sec.~V. 

\section{Method}
The relaxation of the molecular structures, total energy calculations, and electronic structure calculations  are based on a second-order expansion of the Kohn-Sham total energy in density-functional theory with respect to charge density fluctuations.\cite{Frauenheim00}  The zero-th order approach is equivalent to a common standard non-self-consistent tight-binding scheme, while at second order a transparent, parameter-free, and readily calculable expression for generalized Hamiltonian matrix elements can be derived. The resulting DFTB method provides accurate results comparable to the results obtained with \textit{ab-initio} DFT methods or first principle calculations while requiring less computational resources. These are modified by self-consistent redistribution of Mulliken charges. The computations in DFTB+ use a minimal basis set, two center approximation, neglect of three centers and crystal field. The atomic pair potentials are computed at a pre-computation stage using a DFT code. The Slater-Koster integrals are stored (as a function of the inter-atomic distance) in look-up tables which are used during the actual simulation. DFTB+ have been extensively validated in the past years by many applications related to \textit{e.g.}~semiconductor surface reconstructions,\cite{Haugk97,Elstner98,Szuecs03} reaction of organic molecules,\cite{Elstner98,Kruger05}  and biomolecules.\cite{Elstner98,Elstner01} 

Conductance calculations   were done via gDFTB code.\cite{Pecchia04} The gDFTB code is the extension of the DFTB method to include the NEGF formalism combined with the Landauer formalism. The gDFTB code is based on a tight-binding representation of the wave function and a density-functional formulation of the Coulomb and exchange correlation potentials. The conductance at low or zero bias (equilibrium) is obtained from electronic transmission probability $T$ as $G = G_0T(E, V = 0)$, where $G_0 = 2e^2/h$ is the quantum conductance, $e$ is the electron charge, and $h$ is the Plank constant. The electronic transmission probability is related to the retarded/advanced Green function of molecular region $G^{\mathrm{R}/\mathrm{A}}(E)$ and self-energies of the electrodes $\Sigma_{\mathrm{L/R}}(E)$ by Fisher and Lee relation\cite{Fisher81} $T(E,V)= 4\textrm{Tr}[G^\mathrm{R}(E)\textrm{Im}\Sigma_\mathrm{L}(E)G^\mathrm{A}(E)\textrm{Im}\Sigma_\mathrm{R}(E)]$. We emphasize that when calculating the current $I(V) = (2e/h) \int dE[f_\mathrm{L}(E)-f_\mathrm{R}(E)]T(E, V)$, the full voltage-dependent $T(E, V)$ is calculated through self-consistent Poisson-NEGF cycle in gDFTB. Here, $f_{\mathrm{L}/\mathrm{R}}(E)$ are the Fermi functions of the left/right electrode. The minimal basis set used in gDFTB could be responsible for inaccuracies in the calculation of the current. This may result in problems describing transport mediated via the tails of the wave functions. This problem could be solved by using an \textit{ab initio} code with a richer basis set, which would be dramatically increase the computational requirements without adding any qualitative new results to our treatment.

\section{Properties of molecules in gas phase and between Si contacts}
Figure~1(a) shows the two model systems considered in this work. Two different kinds of pairs of linkers are attached to the azobenzene molecules. We refer to these two model systems named as system A  and system B. In each system, two linkers covalently connecting the molecule to silicon contacts are substituted at \textit{meta} and \textit{ortho} positions of the phenyl rings. The large movement of one phenyl ring involved with the \textit{cis-trans} isomerization is unfavorable for the stable operation of molecular switching since such large movement prevents stable coupling between the molecule and Si contacts. However, via the linkers at the \textit{meta} and \textit{ortho} positions unfavorable displacement at molecule-contact interface can be reduced  as shown in Fig.~1(b). Additionally, the change of molecular lengths from one isomeric state to the other is small (see Table~I) so that the molecules do not tilt through the \textit{cis-trans} isomerization. The atoms at contacts do not move throughout, during the \textit{cis-trans} isomerization. Thus, the choice of these positions of substitutions is preferable for the stable operation of molecular switching.

\begin{table}[htb]
\begin{center}
\begin{tabular}{c  r r}
\hline
& \hspace{0.3cm} System A (O-O)&  \hspace{0.3cm} System B (N-N)\\ 
\hline
\textit{cis} & 10.208  \AA & 11.383 \AA \\ 

\textit{trans} & 11.000 \AA & 10.685 \AA \\ 
\hline
\end{tabular}
\end{center}
\caption{Molecular length of two molecular switches in gas phase: The molecular lengths are defined by the atomic pairs in parentheses which make bonds with silicon surfaces at top and bottom. These values are obtained after structure optimization by DFTB+ calculations. For the detail see the text. }
\end{table}

\begin{figure}[ht!]
\begin{center}
\includegraphics[width=7.5cm,clip=true]{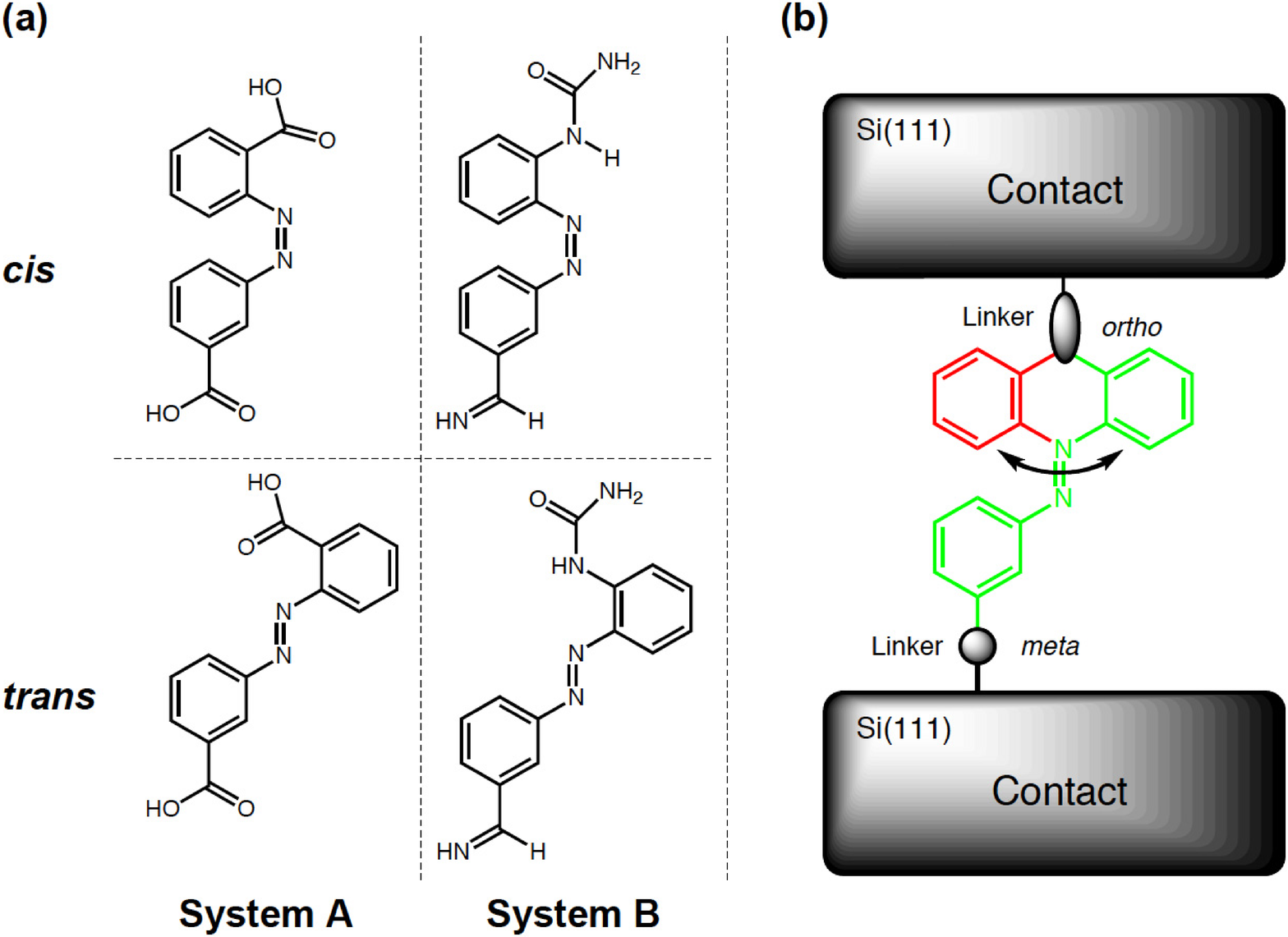}
\end{center}
\caption{\small{ (a) Two azobenzene derivatives modeled in this work and (b) schematic of azobenzene-based molecular switch connected to Si contacts via linkers. The two linkers are substituted at \textit{meta} and \textit{ortho} positions of phenyl rings. These positions of substitutions reduce the movement of atoms at contact involved with \textit{cis-trans} isomerization.}}
\label{DectA}
\end{figure}

%

\subsection{Isolated molecules}
For the determination of stable configuration of the molecules between Si contacts, we first performed a geometry optimization for the  isolated molecules by means of the conjugate-gradient technique. The geometry optimization was carried out until the absolute value of the interatomic force reduced to less than  $10^{-4}$~atomic units. We used the Slater-Koster parameters developed by Elstner \textit{et al.}~for the C, H, N, O atoms\cite{Elstner98} and by Szuecs \textit{et al.}~for the Si atom as basis sets.\cite{Szuecs03} Figure~2(a) and 2(c) show the optimized structures, their frontier orbitals as well as the molecular energy levels of system A and B, respectively. The \textit{trans} isomers are characterized by a planar structure while in each \textit{cis} isomer two phenyl rings are twisted each other of 12.9 degree in system A and B.

Figure~3(a) shows the PES of total energy as a function of the reaction coordinates along the \textit{trans} to \textit{cis} transition for the two model systems in the gas phase. The total energy calculations have shown that \textit{trans} isomers in system A and B are energetically more stable than \textit{cis} counterparts by 0.193~eV and 0.277~eV, respectively. The  \textit{trans} to \textit{cis} potential energy barriers for system A and B are 1.060~eV and 1.433~eV, respectively. These features are similar to the azobenzene molecule in the gas phase where its \textit{trans} isomer is experimentally reported to be more stable than the  \textit{cis} counterpart by 0.6~eV and the potential energy barrier from the \textit{trans} to the \textit{cis} counterpart is 1.0~eV.\cite{Brown75}

\begin{figure*}[ht!]
\begin{center}
\includegraphics[width=15cm,clip=true]{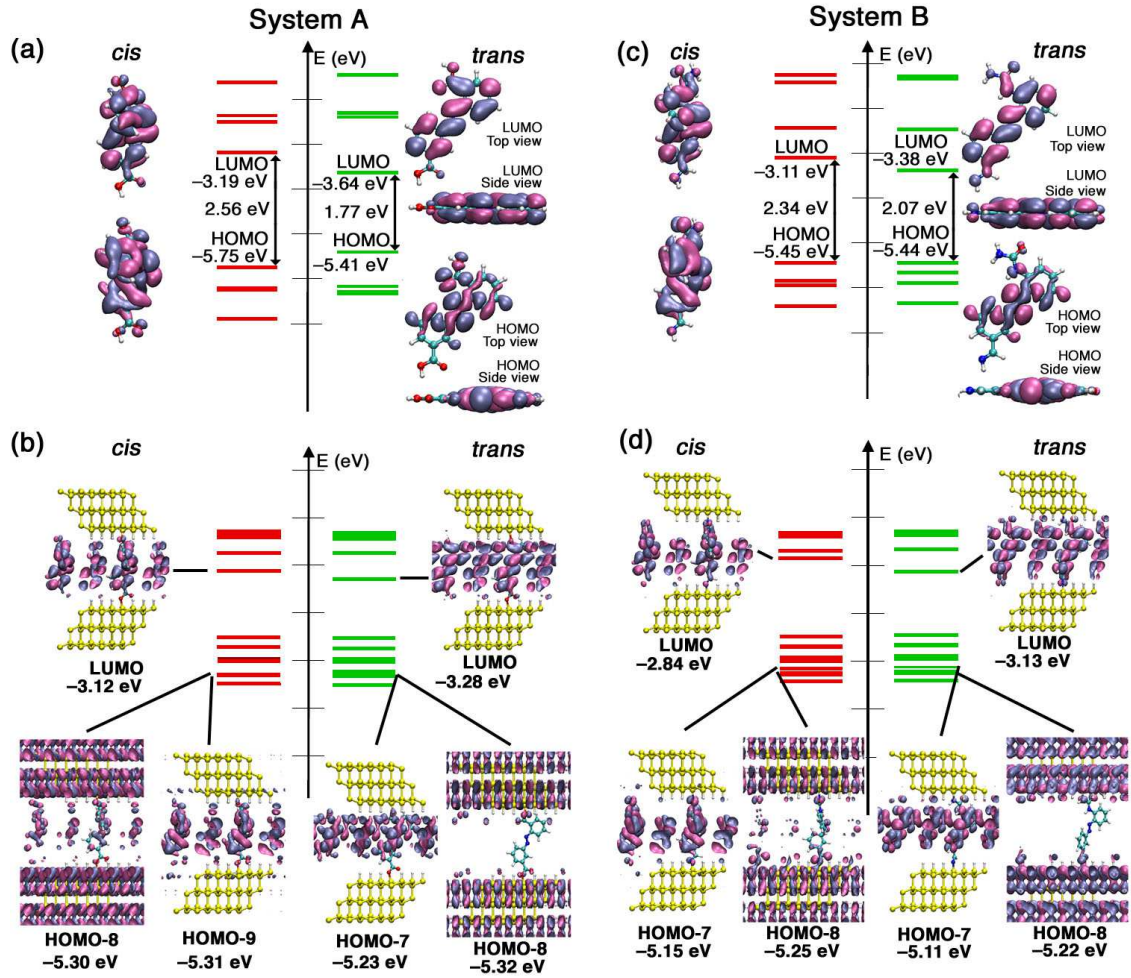}
\end{center}
\caption{\small{Energy level diagram and frontier orbitals of system A in (a) gas phase and (b) between Si contacts. Energy level diagram and frontier orbitals of system B in (c) gas phase and (d) between Si contacts.}}
\label{DectC}
\end{figure*}

%

%
\begin{figure}[ht!]
\begin{center}
\includegraphics[width=8.7cm,clip=true]{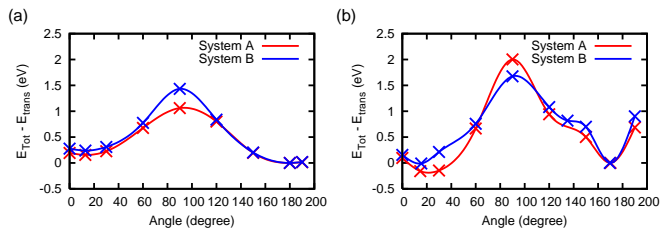}
\end{center}
\caption{\small{Potential energy surface of both systems A and B (a) in gas phase and (b) between Si contacts. The total energies of both systems are plotted as a function of the dihedral angle between the two bonds connecting nitrogen atom and phenyl rings. The curves interpolates points, which are indicated by the crosses.  }}
\label{DectF}
\end{figure}

\subsection{Molecules between Si contacts}
Let us consider the model systems connected between Si contacts. Hereafter we refer to these coupled systems as ``extended molecules". The complexity of the systems leads to a very complicated PES with many metastable states; for simplicity we only consider the vertical arrangements of the molecules with respect to the hydrogen-passivated Si(111) surface. The structure-optimized molecules in the gas phase are covalently connected to hydrogen-passivated Si(111) surfaces and relaxed again using DFTB+ with periodic boundary conditions. The spacing between lateral organic molecules is set large enough to exclude lateral interactions between azobenzene molecules. In each unit cell, the top/bottom silicon contacts are comprised of 54 Si atoms forming three silicon layers and 8 hydrogen atoms covering the Si(111) surface. In order to include the reconstruction of the silicon atoms at the surfaces two silicon layers from the surface are relaxed while the third layers of silicon contacts are fixed, which is required for the subsequent conductance calculations. The Si-Si bond lengths in the fixed layers are set as 2.35~\text{\AA}, which is experimental value for bulk Si crystal. 
Figure~4 shows the total energy of the extended molecules as a function of the separation of two Si surfaces. Figure~5 shows the optimized unit cells of two model systems at the separation of Si surfaces where the total energies of \textit{trans} isomers between Si contacts are their local minimums which are shown in vertical dotted lines in Fig.~4. The PES calculations of extended molecules and subsequent conductance calculations are carried out at these separations.

\begin{table}[htb]
\begin{center}
\begin{tabular}{c  r r }
\hline
&\hspace{0.5cm}System A & \hspace{0.5cm} System B \\ 
\hline
\textit{cis} & $-4.259$~eV & $-4.255$~eV \\  

\textit{trans}& $-4.259$~eV & $-4.253$~eV \\ 
\hline
\end{tabular}
\end{center}
\caption{Fermi energies of two molecular switching junctions. }
\end{table}

\begin{figure}[ht!]
\begin{center}
\includegraphics[width=8.5cm,clip=true]{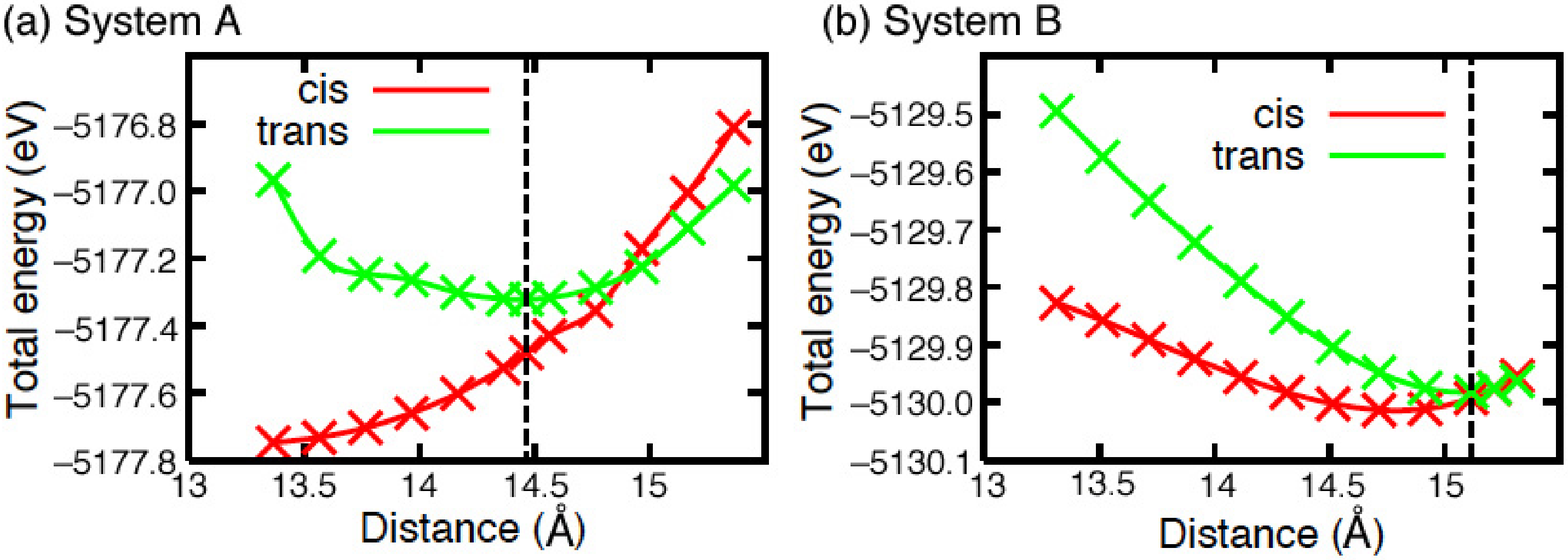}
\end{center}
\caption{\small{ Total energies as a function of distance between two silicon surfaces for two molecular switch junctions. The broken lines indicate the distances where the molecules in \textit{trans} conformation are energetically most stable. All transmission calculations in this article are carried out with these distances.   }}
\label{DectG}
\end{figure}
The PESs of total energies in the extended molecules are shown in Fig.~3(b). The potential energy barriers of extended molecules along the reaction coordinate from \textit{trans} to \textit{cis} isomers in system A and B  are 2.001~eV and  1.675~eV, respectively. The potential energy barriers are increased compared with those in the gas phase (See Fig.~3(a)). This is due to the reduction of the conformational degree of freedom of molecules confined between Si contacts.

\begin{figure}[ht!]
\begin{center}
\includegraphics[width=7.5cm,clip=true]{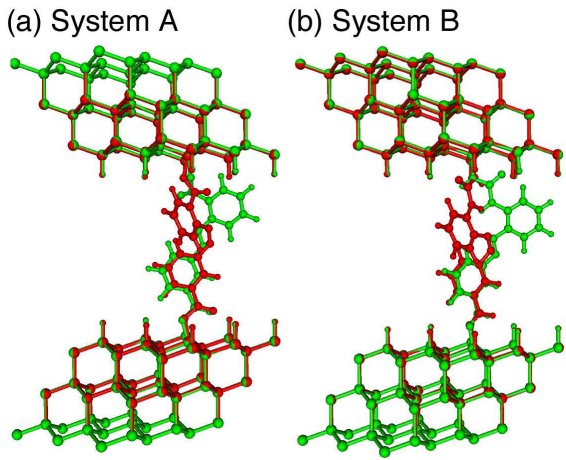}
\end{center}
\caption{\small{Optimized structure (unit cell) of two molecular switch junctions modeled in this work. \textit{cis} (\textit{trans}) conformations are shown in red (green). For comparison, the two isomer conformations are superimposed. Optimizations are carried out using the conjugated-gradient technique with DFTB+ under periodic boundary conditions. The silicon layers at the bottom and top are fixed.}}
\label{DectH}
\end{figure}

 \section{Transport properties}
Once the relaxed geometries have been determined, we first focus on the electronic transmission probabilities of molecular switch junctions at equilibrium for \textit{cis} and \textit{trans} isomer-conformations. The structure-optimized unit cell of the extended molecules is put between semi-infinite Si bulk layers and then transmission calculations are carried out. Figure~6 shows the electronic transmission probabilities at equilibrium. The Fermi energies are near the edge of valence band (for the  Fermi energies, see Table~II). Thus, the conductance in these model systems is driven by hole transport under small bias. Our results show that the \textit{cis} isomers are more conductive  than the \textit{trans} counterparts around the Fermi energies in both of the model systems. The low conductance between the resonant peaks indicates that the transport can only take place through tunneling. The conductance around the Fermi energy in \textit{cis} isomers is higher than that of \textit{trans} counterparts by a few orders of magnitude. This suggests that switching behavior may be expected at low bias. 
The low transmissions around the Fermi energies are due to the weak coupling between the molecules and the Si contacts giving rise to narrow resonant peaks. As shown in the DOS profiles of both system A and B in Fig. 7, the narrow resonant peaks due to the HOMO and LUMO of isolated molecules are located at about $-5.0$ eV and $-3.0$ eV, respectively. The Fermi energies lie in the center of gaps between these narrow resonant peaks due to HOMO and LUMO of isolated molecules. Hence, the DOS on the molecules around the Fermi energies are very low.
\begin{figure}[ht!]
\begin{center}
\includegraphics[width=6.5cm,clip=true]{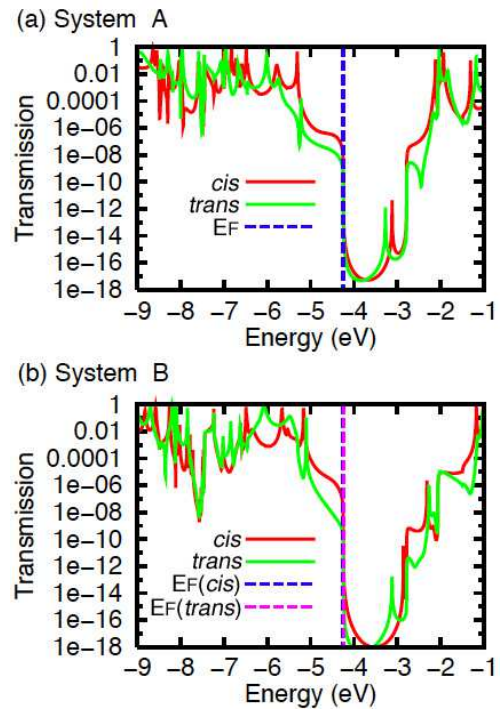}
\end{center}
\caption{\small{ Electronic transmission probabilities for system A and B  at equilibrium. Broken lines show the Fermi energies. In (a), Fermi energies for two isomers are same. Thus, only one broken line is shown. For the detail of Fermi energies, see the Table~II.}}
\label{DectI}
\end{figure}

\begin{figure*}[ht!]
\begin{center}
\includegraphics[width=15.0cm,clip=true]{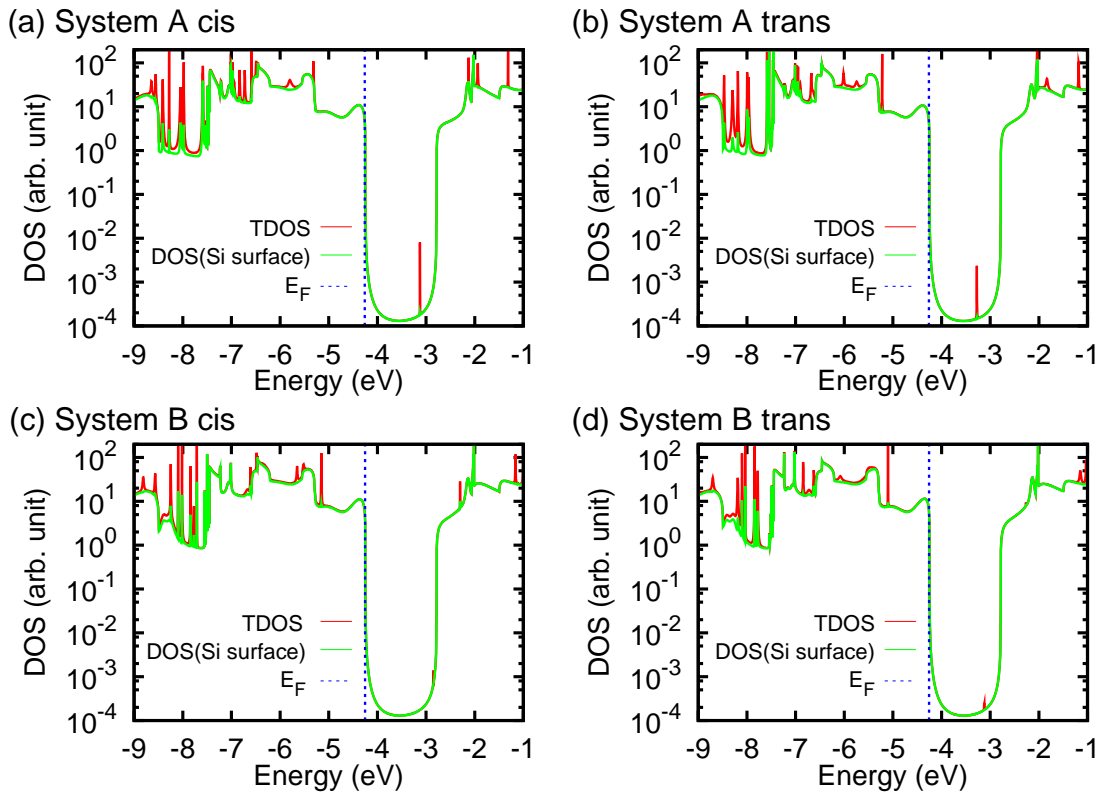}
\end{center}
\caption{\small{ TDOS for system A and B at equilibrium. TDOS are shown in red curves. The contribution from Si surfaces to TDOS is shown in green curves. Broken lines show the Fermi energies.}}
\label{DectIhh}
\end{figure*}

Previous theoretical studies of azobenzene derivatives connected between Au\cite{Zhang04} or CNT\cite{Valle07} electrodes have reported that \textit{cis} isomers are less conductive than their \textit{trans} counterparts since the  $\pi$-conjugation in the \textit{trans} isomers along the molecular framework will be broken by the torsion involved with \textit{cis-trans} isomerizations. However, the same discussion does not hold true in our case  since the HOMO of the \textit{trans} isomers does not consist of $p_z$ orbitals perpendicular to the molecular framework, but  non-binding orbitals mainly localized around nitrogen atoms\cite{Fuchsel06-Sauer08} as shown in Fig.~2(a) and Fig.~2(c). 
The eigenstates consisting of $p_z$ orbitals of the \textit{trans} isomers lie in the HOMO-3 levels.
Thus, these states do not contribute to the transport at low bias.
This counter-intuitive result (\textit{I}(\textit{cis}) $>$ \textit{I}(\textit{trans})) can  also be understood by the analysis of  resulting mixture states of the extended molecular systems as follows. 

Figure 6 shows that the transmission curves for \textit{cis} isomers near the Fermi energies are higher than those of \textit{trans} counterparts, implying higher current through \textit{cis} isomers than the \textit{trans} counterparts  at low bias. The difference between the transmission curves of the \textit{cis} and \textit{trans} isomers near the Fermi energies comes from the difference between the height of the resonant peaks of the \textit{cis} and \textit{trans} isomers around $-5.0$~eV.
The height of the transmission resonant peaks in the two different isomer-conformations can be interpreted from the molecular orbitals of the extended molecules. The HOMOs of the \textit{cis} isomers interact with surface states of the Si contacts and their orbitals hybridize. The resulting mixed states,  corresponding to the HOMO-8 of systems A in Fig.~2(b) and of system B in Fig.~2(b), have a non-negligible orbital distribution bridging across the Si states. These states can contribute to transport as shown in the resonance peaks (in red) of system A at $-5.31$~eV in Fig.~6(a), of system B at $-5.17$~eV in Fig.~6(b). 

On the other hand, the HOMOs of the \textit{trans} isomers  have neither orbital amplitude around the linkers nor interact strongly with the surface states of the Si contacts. This leads to less orbital hybridization as can be seen for  HOMO-8 of Figures 2(b) and (d).  These states do not contribute to transport as shown by the sharper and smaller resonant peaks (in green) for system A at $-5.22$~eV in Fig.~6(a) and for system B at $-5.10$~eV in Fig.~6(b). Despite the delocalized molecular orbitals along the molecular framework, the LUMOs of both the \textit{trans} and \textit{cis} isomers in all model systems do not interact with the surface state of the Si surface. This is because  the  energetic difference between  LUMO of both \textit{cis} and \textit{trans} isomers and conduction state of the Si surface (the conduction band is above $-2.70$~eV and valence band is below  $-4.51$~eV)  is large. Additionally, there is no state available in the silicon contacts  on the energy around LUMOs of both the \textit{trans} and \textit{cis} isomers in both of the model systems. Thus, these states lead to much lower peaks around $-3.0$~eV in transmission spectra as shown in Fig.~6.

We have to pay attention not only to the $\pi$-conjugation but also to the interaction between the molecule and the contacts. These results support the conclusion that the electronic structure of both molecule and contact are crucial in determining the transport properties at low bias.

\begin{figure}[ht!]
\begin{center}
\includegraphics[width=8.5cm,clip=true]{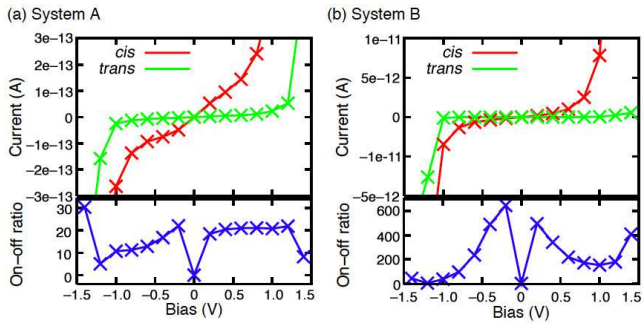}
\end{center}
\caption{\small{\textit{I-V} characteristics of molecular switch junctions in two different conformations: The bottom panels show the on-off ratios (the ratio of current between \textit{cis} and \textit{trans} isomer). In all cases, the \textit{cis} conformations (red) are conductive than \textit{trans} counterparts (green).}}
\label{DectJ}
\end{figure}
In order to confirm the possible switching behavior already suggested by the analysis of transmission probabilities at equilibrium, we have calculated the current-voltage (\textit{I-V}) characteristics for the two different isomer-conformations of the two model systems. The \textit{I-V} characteristics for \textit{cis} and \textit{trans} conformations for the two model systems are shown in Fig.~8. The charge transport in \textit{trans} isomers is strongly suppressed compared to that in the \textit{cis} counterparts for both of the two model systems. If the bias voltage is less than 1.0~V, the resonant peaks do not enter Fermi bias window so that the \textit{I-V} curves are smooth within the bias window. Therefore, we can expect constant on-off ratios as a response to the change or fluctuation of bias voltage, which are preferable for the robust control of the molecular switching. The bottom panels in Fig.~8(a) and (b) show the on-off ratios (the ratio of current in \textit{cis} isomer to \textit{trans} counterpart). The on-off ratio reaches up to 647 for system B at 0.2~V (bottom panel of Fig.~8(b)). Although the on-off ratios decrease with higher bias voltage, all molecular switches recorded on-off ratios over 10 within 1.0~V. Higher on-off ratios for lower bias voltage are advantageous both for power saving and for the suppression of unexpected breakdown triggered at higher bias voltage.

\begin{figure}[ht!]
\begin{center}
\includegraphics[width=8.5cm,clip=true]{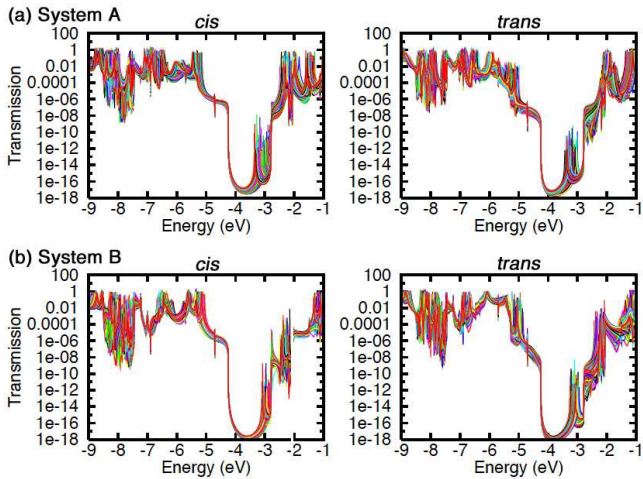}
\end{center}
\caption{\small{Time dependent fluctuations of transmission spectra along MD pathways at equilibrium for two molecular switch junctions in Fig.~1. In each graph, 100 transmission spectra are superimposed. The time interval of 100 curves is 100~fs.}}
\label{DectK}
\end{figure}
   
 \subsection{Time dependent fluctuation of conductance}
For the reliable operation of molecular switching one needs to investigate the stability of the model systems between Si contacts and the fluctuation of the current through them in ambient conditions. For this purpose, we have performed MD simulations of the model systems between Si contacts and calculated the conductance of the two different isomer-conformations along the MD pathways. The calculation setting was as follows. We used the Verlet algorithm for the MD simulation in DFTB+ with periodic boundary conditions. We used the same SK parameters used for the structure optimization. The MD duration is set for 10~ps, \textit{i.e.}, 10,000 steps at interval of 1.0~fs. The nuclear and electronic temperatures are set as 300~K. After performing MD simulations 
the coordinates in every 100 steps are extracted from MD trajectories. Finally we have repeated conductance calculations for collected 100 coordinates using the gDFTB code both with and without applied bias voltage. We then analyzed the fluctuation of the transmission probabilities and \textit{I-V} curves along the MD pathways.

\begin{figure}[ht!]
\begin{center}
\includegraphics[width=8.7cm,clip=true]{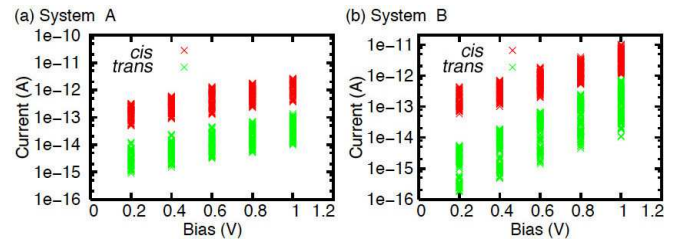}
\end{center}
\caption{\small{The dispersion of currents along MD pathways as a function of bias voltage. In each bias voltage 100 values of the current in \textit{cis} isomers (\textit{trans} isomers) along the MD pathways are plotted in red (green) crosses.}}
\label{DectL}
\end{figure}

The transmission probabilities along the MD pathways are plotted in Fig.~9. In each isomer 100 transmission spectra are plotted. We can see that the main features of the transmission spectra in Fig.~6 remain unchanged. Hence, we can expect that the fluctuation of current flow under ambient conditions is small. The dispersion of the current along the MD pathways as a function of bias voltage is plotted in Fig.~10. The dispersion of the current in the two isomers is distinguishable, with the \textit{cis} isomers always being more conductive than the \textit{trans} counterparts. The currents in the two isomers as a function of time under bias (0.2~V) in Fig.~11  also support this conclusion. The on-off ratio is over 15 (the minimum value is 15.5 at 4.7~ps in Fig.~11(b)) for all model systems. Therefore, the molecular switch junctions modeled on this study may be expected to work as switching components in ambient conditions. 


  %
\begin{figure}[ht!]
\begin{center}
\includegraphics[width=8.7cm,clip=true]{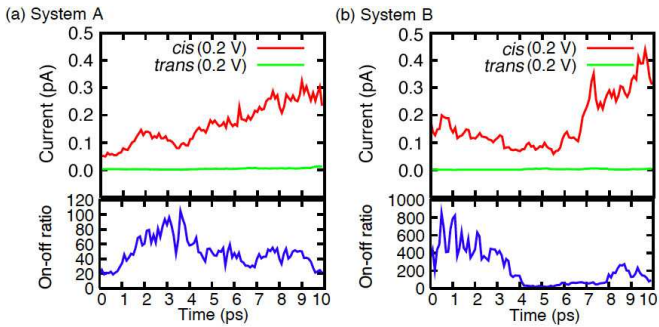}
\end{center}
\caption{\small{Time dependent fluctuation of current (top) and on-off ratio (bottom) along MD pathways in molecular switch junctions. }}
\label{DectM}
\end{figure}

Finally we would like to comment on the possible experimental realization of molecular conformational switching device consisting of azobenzene derivative and Si contacts. The attachment of aromatic compounds to the Si surface will be possible using the methodology developed by Murata \textit{et al.}\cite{Murata,ESPMI} The realization of two terminal Si/molecule/Si devices by covering of the functionalized Si surfaces from the top with a Si contact might present a formidable challenge due to the intrinsic asymmetry in molecular junctions. 
However, first attempts to such a difficult task proved already successfully.\cite{Nature-PSS}
For further understanding and control of the molecular switching, systematic studies investigating the effect of contacts,\cite{Fagas} internal scattering effect,\cite{Nozaki} surface roughness, and impurities\cite{Lucio} will be required. The spacing between two Si surfaces will also be an important factor since it will influence the PES, molecular and electronic structure, and resulting conductivity of the molecule embedded between Si contacts. Concerning the design of azobenzene derivatives, further investigation is required to determine which linkers and functional groups lower the PES barrier for \textit{cis-trans} isomerization. One possibility to achieve it is to increase the flexibility of the molecule by using long linkers which should not, however, increase the fluctuation of current due to the flexibility of the molecules.

\section{Conclusion}
In summary, we have modeled two molecular switch junctions consisting of azobenzene derivatives and silicon contacts and explored their potential as molecular switches by the analysis of the energetics, PES, and transport properties in the two different isomer-conformations:~\textit{cis} and \textit{trans}. 
Modeling of the molecular switches demonstrated that the attachment of linkers at the \textit{meta} and \textit{ortho} positions could reduce the steric repulsion between the molecules and silicon contacts during \textit{cis-trans} isomerization to a value which is reasonable for the robust operation of molecular switching. 
Our numerical results show that \textit{cis} isomers are more conductive than the \textit{trans} counterparts inside the bias window for both model systems. \textit{I-V} characteristics also show the same trends. Although the  $\pi$-conjugation through the molecular framework is reduced due to the torsion of two phenyl rings, the \textit{cis} isomers interact much more strongly with the surface states of the silicon contacts than their \textit{trans} counterparts, resulting in larger conductivities. Additionally, the calculation of currents through the two model systems along the MD pathways in ambient conditions revealed that the currents in \textit{cis} isomers are always significantly larger than \textit{trans} counterparts leading to high on-off ratio. Importantly, the dispersion of the current as a function of time is small enough to distinguish two different isomer-conformations. Thus, azobenzene derivatives between Si contacts are promising for applications as robust molecular switching components or single molecular data storage devices.

\section*{Acknowledgments}
This work has been partially funded by the Volkswagen Foundation and by the WCU (World Class University) program through the Korea Science and Engineering Foundation funded by the Ministry of Education, Science and Technology (Project No. R31-2008-000-10100-0). We acknowledge the Center for Information Services and High Performance Computing (ZIH) at the Dresden University of Technology for computational resources. We thank Rafael Guti\'{e}rrez, Haldun Sevincli, Florian Pump and Cormac Toher for fruitful discussions.

\end{document}